\documentclass[usenatbib]{mn2e}
\usepackage{natbibmnfix,graphicx,times}

\newcommand{\Mpc}{\mbox{ Mpc}}

\newcommand{\erg}{\mbox{ erg}}
\newcommand{\keV}{\mbox{ keV}}

\newcommand{\kel}{\mbox{ K}}
\newcommand{\mkel}{\mbox{ mK}}

\newcommand{\ergsec}{\mbox{ erg s$^{-1}$}}

\newcommand{\MHz}{\mbox{ MHz}}

\newcommand{\Msun}{\mbox{ M$_\odot$}}

\newcommand{\sfr}{\mbox{ M$_\odot$ yr$^{-1}$}}

\newcommand{\hunits}{\mbox{ km s$^{-1}$ Mpc$^{-1}$}}

\newcommand{\recunits}{\mbox{ cm$^{3}$ s$^{-1}$}}

\newcommand{\Junits}{\mbox{ cm$^{-2}$ s$^{-1}$ Hz$^{-1}$ sr$^{-1}$}}

\newcommand{\bxhi}{\bar{x}_{\rm HI}}
\newcommand{\xhi}{x_{\rm HI}}

\newcommand{\bxion}{\bar{x}_i}
\newcommand{\xion}{x_i}
\newcommand{\dtb}{\delta T_b}
\newcommand{\bdtb}{\bar{\delta T}_b}

\newcommand{\htwo}{HII }
\newcommand{\mmin}{m_{\rm min}}
\newcommand{\fcoll}{f_{\rm coll}}
\newcommand{\fesc}{f_{\rm esc}}
\newcommand{\lya}{Ly$\alpha$ }
\newcommand{\lyn}{Ly$n$ }

\newcommand{\deriv}{d}
\def\VEV#1{\left\langle #1\right\rangle} % This is \VEV{x} => <x>

\newcommand{\apj}{ApJ \ }
\newcommand{\apjl}{ApJ \ }
\newcommand{\apjs}{ApJS \ }
\newcommand{\aap}{A\&A \ }
\newcommand{\aj}{AJ \ }
\newcommand{\mnras}{MNRAS \ }

\newcommand{\physrep}{Physics Reports \ }

\newcommand{\nat}{Nature \ }

\title[The Global 21 cm Background]{The global 21 centimeter background from high redshifts}

\author[S.~R. Furlanetto]{Steven R.  Furlanetto\thanks{Email:sfurlane@tapir.caltech.edu} \\
Yale Center for Astronomy and Astrophysics, Yale University, 260 Whitney Avenue, New Haven, CT 06520-8121}

\begin{document}

\maketitle

\begin{abstract}

We consider the evolution of the sky-averaged 21 cm background during the early phases of structure formation.  Using simple analytic models, we calculate the thermal and ionization histories, assuming that stellar photons dominate the radiation background.  The resulting 21 cm spectra can constrain the properties of the first generations of stars and quasars.  If Population II stars dominate, \lya coupling renders the IGM visible before it is heated by X-rays and long before reionization.  Thus the 21 cm background has a strong absorption epoch followed by weaker emission that fades during reionization.  The harder spectra of very massive Population III stars compress these transitions into a shorter time interval and decreases the signal amplitude.  However, the reionization epoch will remain visible except in extreme cases.  The global 21 cm signal will be challenging to observe because of astronomical foregrounds, but it offers an exciting opportunity to study the first sources of light.  It also fixes the overall amplitude of the fluctuating background whose detection is a major goal of several next-generation low-frequency radio interferometers.

\end{abstract}


\section{Introduction} \label{intro}

The earliest generations of cosmic structures have profound effects on the Universe around them and on later generations of stars and galaxies.  These sources ``reionize" the intergalactic medium (IGM), enrich galaxies (and the IGM) with the first heavy elements, and seed larger galaxies.  

Observations now paint a fascinating -- but confusing -- picture of their properties.  Unfortunately, observing ``typical" galaxies at $z \ga 8$ is beyond the capabilities of existing instruments (though see \citealt{bouwens04}), so current constraints use reionization as an indirect proxy.  The rapid evolution of the mean \lya optical depth in quasar absorption spectra seems to indicate that reionization is ending at $z \sim 6$ \citep{fan01, fan06}.  Corroborating evidence comes from the proximity zones around these quasars, which also imply rapid evolution in the mean ionized fraction $\bxion$ \citep{wyithe04-prox, mesinger04, wyithe05-prox, fan06}, as well as large-scale fluctuations in the transmission.  But both are difficult to interpret:  the latter, for example, \emph{may} also be characteristic of reionization \citep{wyithe06-var, fan06}, but so far the variations are also consistent with density fluctuations \citep{lidz06}.  On the other hand, large-scale polarization of the cosmic microwave background (CMB) observed by \emph{WMAP} implies an early beginning to reionization at $z \ga 10$, albeit with large uncertainties \citep{page06, spergel06}.  How these puzzle pieces fit together -- and, more fundamentally, how luminous sources evolve, how feedback processes (such as metal enrichment, photoionization, or other radiation backgrounds) influence them, and how small-scale structure evolves in the IGM-- remain mysterious \citep{wyithe03, cen03, wyithe03-letter, cen03-letter, haiman03, somerville03, fukugita03, onken04, furl05-double, choudhury05, iliev05}.  Unfortunately, identifying the relevant feedback processes and uncovering the source properties are impossible with existing data.

One promising method to disentangle these degeneracies is the 21 cm transition of neutral hydrogen (see \citealt{furl06-review} for a review).  Before reionization, the neutral IGM gas should be visible against the CMB through this transition \citep{field58, sunyaev72, hogan79}.  In principle, we can use it to map out the growth of the cosmic web and \htwo regions at $z \ga 6$ \citep{scott90, madau97, zald04}.  The key advantage of the 21 cm transition is that it is a line:  hence every observed frequency corresponds to a different distance, and we can isolate each phase in the IGM's history.  

Of the many applications of the 21 cm line, the simplest (at least conceptually) is to measure the global background as a function of frequency.  This would allow us to measure $\bxion(z)$ and, at earlier redshifts, the growth of soft-ultraviolet (UV) and X-ray radiation backgrounds, which are difficult to constrain in any other way.  Three quantities are relevant: the thermal history, the \lya background (both of which determine the spin temperature $T_S$ of the IGM), and the ionization history (which determines when the signal fades).  Unfortunately, this global background is likely to be difficult to measure because of contamination by Galactic synchrotron radiation (see \S \ref{disc}).  As a result, it has received little theoretical attention, and most existing work has simply argued that $T_S \gg T_\gamma$ (where $T_\gamma$ is the CMB temperature) during reionization (e.g., \citealt{madau97, ciardi03-21cm, chen04}).  This is unfortunate because a number of efforts are nevertheless underway to observe it (and the challenges are not necessarily much greater than observing fluctuations in the background).  Indirect techniques may also allow us to measure the background \citep{barkana05-ts, cooray05-glob}.  And, of course, the global signal sets the overall amplitude of fluctuations, so those should be considered within this context.  One exception is \citet{sethi05}, who studied the evolution of the global background in a limited set of models that did not fully explore the many uncertainties in the properties of the first sources.  

The purpose of this work is to compute the high-redshift 21 cm signal and to identify those properties that can be measured from the mean background.  Because we know so little about the first sources, it is of course impossible to make robust predictions for the background.  We will instead parameterize the relevant processes and consider how they affect the mean signal.  We first give some background on the 21 cm transition in \S \ref{21cm}.  We then describe processes relevant to the thermal history of the IGM in \S \ref{therm}, to the spin temperature in \S \ref{tshist}, and to the ionization history in \S \ref{ionhist}.  We then describe some general consequences for the 21 cm history in \S \ref{critpt} and examine some specific models in \S \ref{glob-theor}.  We conclude in \S \ref{disc}.

In our numerical calculations, we assume a cosmology with $\Omega_m=0.26$, $\Omega_\Lambda=0.74$, $\Omega_b=0.044$, $H=100 h \hunits$ (with $h=0.74$), $n=0.95$, and $\sigma_8=0.8$, consistent with the most recent measurements \citep{spergel06}, although we have increased $\sigma_8$ from the best-fit \emph{WMAP} value in order to improve agreement with weak-lensing data.

\section{The 21 cm Transition} \label{21cm}

We review the relevant characteristics of the 21 cm transition here; we refer the interested reader to \citet{furl06-review} for a more comprehensive discussion.  The 21 cm brightness temperature (relative to the CMB) of a patch of the IGM is
\begin{eqnarray}
\dtb & = & 27 \, \xhi \, (1 + \delta) \, \left( \frac{\Omega_b h^2}{0.023} \right) \left( \frac{0.15}{\Omega_m h^2} \, \frac{1+z}{10} \right)^{1/2} \nonumber \\ 
& & \times \left( \frac{T_S - T_\gamma}{T_S} \right) \mkel,
\label{eq:dtb}
\end{eqnarray}
where $\delta$ is the fractional overdensity, $\xhi = 1 - \xion$ is the neutral fraction, and $x_i$ is the ionized fraction.  Note that the patch will appear in absorption if $T_S < T_\gamma$ and emission otherwise.  Here we have assumed that the patch expands uniformly with the Hubble flow; radial peculiar velocities also affect $\dtb$ by changing the mapping from distance to frequency \citep{bharadwaj04-vel,barkana05-vel}.  However, these anisotropic fluctuations do not affect the mean signal, so we will ignore them.  We will use $\bxion$ to denote the globally averaged ionized fraction; the temperatures are likely to be much more uniform at most times, and we will implicitly treat them as independent of position (though see \citealt{barkana05-ts}).

The spin temperature $T_S$ is determined by competition between three processes:  scattering of CMB photons, collisions, and scattering of Ly$\alpha$ photons \citep{wouthuysen52,field58}.  In equilibrium \citep{hirata05},
\begin{equation}
T_S^{-1} = \frac{T_\gamma^{-1} + \tilde{x}_\alpha \tilde{T}_c^{-1} + x_c T_K^{-1}}{1 + \tilde{x}_\alpha + x_c}.
\label{eq:tsdefn}
\end{equation}
Here $x_c \propto n$ is the collisional coupling coefficient for H--H interactions \citep{zygelman05} and H--e$^-$ collisions \citep{smith66, liszt01}.   The middle term describes the Wouthuysen-Field effect, in which absorption and re-emission of Ly$\alpha$ photons mixes the hyperfine states.  The coupling coefficient is \citep{chen04, hirata05}
\begin{equation}
\tilde{x}_\alpha = 1.81 \times 10^{11} (1+z)^{-1} \tilde{S}_\alpha J_\alpha,
\label{eq:xalpha}
\end{equation}
where $\tilde{S}_\alpha$ is a factor of order unity describing the detailed atomic physics of the scattering process and $J_\alpha$ is the background flux at the Ly$\alpha$ frequency in units $\Junits$; the Wouthuysen-Field effect becomes efficient when there is about one photon per baryon near this frequency.  It couples $T_S$ to an effective color temperature $\tilde{T}_c$ (in most circumstances, $\tilde{T}_c \approx T_K$; \citealt{field59-ts}).  We use the numerical fits of \citet{hirata05} for $\tilde{S}_\alpha$ and $\tilde{T}_c$.

\section{The Thermal History} \label{therm}

To compute the evolution of $\dtb$, we therefore need $T_K(z)$, $J_\alpha(z)$, and $\bxion(z)$.  In the next three sections, we will examine each in turn, beginning with the kinetic temperature.  The evolution equation is
\begin{equation}
\frac{\deriv T_K}{\deriv t} = -2 H(z) T_K + \frac{2}{3} \sum_i
\frac{\epsilon_i}{k_B n},
\label{eq:tkevol}
\end{equation}
where the first term on the right hand side is the $p \, \deriv V$ work from the expanding universe and $\epsilon_i$ is the energy input into the gas per second per unit (physical) volume through process $i$.  We include a number of possible input mechanisms.

\subsection{Compton Heating} \label{comp}

At high redshifts, Compton scattering between CMB photons and the residual free electrons in the IGM dominates.  The heating rate is (e.g., \citealt{seager00})
\begin{equation}
\frac{2}{3} \, \frac{\epsilon_{\rm comp}}{k_B n} = \frac{\bxion}{1 + f_{\rm He} + 
  \bxion} \, \frac{8 \sigma_T u_\gamma}{3 m_e c} \, (T_\gamma - T_K),
\label{eq:tcomp}
\end{equation}
where $f_{\rm He}$ is the helium fraction (by number), $u_\gamma \propto T_\gamma^4$ is the energy density of the CMB, and $\sigma_T$ is the Thomson cross-section.  The first factor occurs
because CMB photons scatter off free electrons, but the heat must be shared with all the particles.  The efficiency of Compton heating decreases with cosmic time as electrons recombine and $u_\gamma$ falls; the IGM thermally decouples from the CMB at \citep{peebles93}
\begin{equation}
1 + z_{\rm dec} \approx 150 (\Omega_b h^2/0.023)^{2/5}.
\label{eq:zdec}
\end{equation}
Below this point, $T_K \propto (1+z)^2$, as appropriate for a thermally isolated expanding gas.

\subsection{X-ray Heating} \label{xrayheat}

The thermal evolution becomes much more complex (and uncertain) once the first luminous sources appear.  X-rays from galaxies and quasars likely provide most of the heat for the low-density IGM.  Given our large uncertainties about the nature of high-redshift objects, we will take a simple parameterized approach calibrated to nearby starburst galaxies.  These objects show a clear correlation between star formation rate (as measured by the total infrared luminosity) and hard X-ray (2--10 keV) luminosity \citep{grimm03,ranalli03,gilfanov04}:
\begin{equation}
L_X = 3.4 \times 10^{40} f_X \left( \frac{{\rm SFR}}{1 \sfr} \right) \ergsec,
\label{eq:sfrxray}
\end{equation}
where $f_X$ is a correction factor accounting for the (unknown) differences at high redshifts.  Note that we have transformed to the total ($>0.2 \keV$) X-ray luminosity assuming a spectral index
$\alpha=1.5$ (where $L_\nu \propto \nu^{-\alpha}$) in this range \citep{rephaeli95}.  This is only an average index, of course, but errors can be subsumed into the unknown normalization $f_X$.

There is of course no rigorous justification for extrapolating this relationship to high redshifts, but it is certainly reasonable.  Massive X-ray binaries produce the bulk of high-energy emission in nearby starbursts.  Such systems form when the first massive stars die, only a few million years after star formation commences.  So they should also be present in high-redshift galaxies \citep{glover03}.  Of course, their abundance must change with the metallicity and initial mass function (IMF):  if, for example, the IMF is top-heavy, $f_X$ could be significantly larger than unity.  But for the moment we can only speculate about the magnitude of such evolutionary effects.

A second source is inverse-Compton scattering from relativistic electrons accelerated in the supernova remnants.  In the nearby universe, only powerful starbursts have strong enough radiation fields for this to be competitive with synchrotron cooling; however, the CMB has $u_\gamma \propto (1+z)^4$, so  inverse-Compton emission may dominate over synchrotron emission in {\em all} star-forming environments at high redshift \citep{oh01}.  We can make a simple estimate of the extra luminosity contributed by these fast electrons.  Suppose that a fraction $\epsilon_e$ of the total supernova energy $E_{\rm SN}$ is used to accelerate electrons, and that a fraction $f_e \sim 0.5$ of this energy is emitted in X-ray photons (this fraction is determined by cooling within the shock; \citealt{furl04-sh}).  We also let $\nu_{\rm SN}$ be the number of supernovae per unit mass of star formation.  Then
\begin{eqnarray}
L_X^{\rm SN} & = & 1.6 \times 10^{40} \, f_e \, \left( \frac{\epsilon_e}{0.05} \, \frac{\nu_{\rm SN}}{0.01 \Msun^{-1}} \, \frac{E_{\rm SN}}{10^{51} \erg} \right) \nonumber \\ 
& & \times \left( \frac{{\rm SFR}}{1 \sfr} \right) \ergsec,
\label{eq:snxray}
\end{eqnarray}
where we have substituted $\epsilon_e \sim 0.05$ as appears to be appropriate for nearby supernova remnants (e.g., \citealt{koyama95}).  Thus, if a large fraction of the electron energy is emitted in X-ray photons, inverse-Compton emission from fast electrons can contribute a non-negligible -- though not necessarily dominant -- fraction of the total emission.  Conveniently, $f_X \sim 1$ still seems reasonable.  

Extremely massive ($\ga 300 \Msun$) Pop III stars\footnote{Note that we will use ``Pop III" to refer exclusively to this very massive case.  If Pop III stars have more or less normal initial mass functions, they produce histories close to our Pop II case.  We use the massive stars only for contrast with the ``standard" histories and do not mean to pass judgment on the relative merits of the different initial mass functions.} could have a wildly different normalization, of course:  each pair-instability supernova releases $\sim 10^{52}$--$10^{53} \erg$, much more than normal supernovae \citep{heger02}.  The fraction of X-ray binaries may also be significantly larger, or the remnants may form ``miniquasars."  Moreover, these stars also produce soft X-rays directly, because they have surface temperatures $T_{\rm eff} \sim 10^5 \kel$ \citep{bromm01-vms, schaerer02}.  To a first approximation, the stellar spectra are simple blackbodies, in which case $\sim 0.3\%$ of the total stellar luminosity is emitted in photons with $E>0.1 \keV$.  This would imply $f_X \gg 1$.  However, photons near $\sim 0.1 \keV$ are absorbed relatively quickly, and they affect only the region within $\la 0.5 \Mpc$ of galaxies \citep{sethi05, cen06}.  Thus it is the harder photons with which we are primarily concerned, and these come from nonthermal sources: at $T_{\rm eff} = 10^5 \kel$, only $\sim 2 \times 10^{-7}$ of the Pop III stellar luminosity lies in photons with $E>0.2 \keV$.  

Because X-ray heating is thus intrinsically non-uniform, we must develop some approximation for our global treatment -- in the simplest terms, we must specify the photon energy range that we assume can penetrate into the diffuse IGM.  Equation (\ref{eq:sfrxray}) included all photons with $E>0.2 \keV$.  In actuality, all photons with $E<E_{\rm thick}=1.5 \, \bxhi^{1/3} [(1+z)/10]^{1/2} \keV$ are absorbed within a Hubble length \citep{oh01}, so photons near the bottom edge of our range will be absorbed over a relatively small -- but by no means negligible -- volume.  By including soft photons, we are attempting to balance the highly-penetrating hard X-rays with the more localized soft X-rays that likely carry more total energy.  Of course, this fluctuating component (e.g., \citealt{madau04}) will lead to non-uniform heating \citep{sethi05}.  We defer discussion of these fluctuations to future work; our concern here is with understanding the zeroth-order global evolution.  Fortunately, uncertainty in the overall effect of inhomogeneity can be subsumed into our much larger uncertainty in $f_X$ so long as the spectrum is a reasonably shallow power law (as appropriate for nonthermal sources).  For example,  \citet{glover03} used a method much like our own but took $f_X \sim 0.2$ because they calibrated to the luminosity in the 2--10 keV band, which implicitly assumes that soft photons are absorbed inside or near their source galaxies.

X-rays deposit energy in the IGM by photoionizing hydrogen and helium; the hot ``primary'' electron then distributes its energy through three channels: (1) collisional ionization, producing more secondary electrons, (2) collisional excitation of He, which produces a photon capable of ionizing H, and (3) Coulomb collisions with thermal electrons.  The relative cross-sections of these processes determine the fraction of X-ray energy going to heating ($f_{X,h}$) and ionization ($f_{X,{\rm ion}}$); clearly they depend on both $\bxion$ and the initial photon energy.  For our calculations, we will use the high energy limit ($E \gg 0.1 \keV$) of \citet{shull85}, who fit their results to
\begin{eqnarray}
f_{X,h} & = & C_1 [ 1 - (1-\xion^{a_1})^{b_1} ] \nonumber \\
f_{X,{\rm ion}} & = & C_2 (1-\xion^{a_2})^{b_2}.
\label{eq:shullfit}
\end{eqnarray}
Here $C_1=0.9971$, $a_1=0.2663$, $b_1=1.3163$, $C_2=0.3908$, $a_2=0.4092$, and $b_2=1.7592$; the fits are accurate to $\la 2\%$ in the range of interest.  Although not as good for softer photons, the qualitative behavior is similar:  when $\bxion$ is small, most of the energy is deposited in ionizations ($f_{X,h} \approx 0.2$ for $E=3 \keV$ and $\bxion=10^{-3}$), but $f_{X,h} \rightarrow 1$ as $\bxion \rightarrow 1$, because collisions with free electrons become more important.\footnote{Note that here $\xion$ refers to the ionized fraction in the predominantly neutral gas (outside of \htwo regions) whose temperature we care about.  In the models below, we track this separately from the usual ionized fraction, which instead describes the filling factor of \htwo regions, by assuming that only X-rays ionize this mostly neutral gas.}

We are now in a position to estimate the heating rate from X-rays in the early universe.  We first assume that the star formation rate is proportional to the rate at which matter collapses into galaxies, which we assume correspond to dark matter halos with $m > \mmin$, where $\mmin$ is a (redshift-dependent) minimum mass.  We will typically take $\mmin=m_4$, where $m_4$ has a virial temperature $T_{\rm vir}=10^4 \kel$ (the threshold for atomic cooling; \citealt{barkana01}).   We define $\fcoll$ to be the fraction of matter in collapsed halos with $m > \mmin$; for the \citet{press74} mass function, it takes the simple analytic form
\begin{equation}
\fcoll = {\rm erfc} \left[ \frac{\delta_c(z)}{\sqrt{2} \sigma(\mmin) }
  \right],
\label{eq:fcollps}
\end{equation}
where $\delta_c$ is the linear overdensity at virialization and $\sigma^2(m)$ is the variance of the density field when smoothed on scale $m$.  By equation (\ref{eq:sfrxray}), the X-ray emissivity is proportional to the star formation rate and hence to $\deriv \fcoll/\deriv t$; the result is
\begin{equation}
\frac{2}{3} \, \frac{\epsilon_{X}}{k_B n H(z)} = 10^3 \kel \ f_X \, \left(
\frac{f_\star}{0.1} \, \frac{f_{X,h}}{0.2} \, \frac{\deriv
  \fcoll/\deriv z}{0.01} \, \frac{1+z}{10} \right),
\label{eq:xrayemiss}
\end{equation}
where $f_\star$ is the star formation efficiency.  We immediately see that X-ray heating should be quite rapid, provided that $f_X \ga 0.1$.

Of course, our assumption that $\epsilon_X \propto \deriv \fcoll/\deriv t$ may be wrong because (for example) some exotic process produces X-rays, or because quasars are common, or simply because $f_\star$ is not a constant.  However, unless the process of interest is uncorrelated with galaxy formation,  it will not differ dramatically from that we have discussed.  On the other hand, quasars could strongly increase the normalization of $f_X$.  We have chosen to ignore this possibility because extrapolating the known population of bright quasars to higher redshifts yields an X-ray background with negligible effects on the IGM \citep{venkatesan01}.  A number of models posit an entirely new population of ``miniquasars" (e.g., \citealt{madau04, ricotti05}), but we will conservatively ignore them.  We briefly discuss how such miniquasars could affect our conclusions in \S \ref{disc}.

\subsection{\lya Heating} \label{lyaheat}

As photons redshift into the Lyman-series resonances, they scatter off of neutral hydrogen atoms in the IGM and deposit energy in the gas through atomic recoil.  While initially thought to be an important heat source \citep{madau97}, \citet{chen04} and \citet{rybicki06} showed that the energy deposition rate in repeated \lya scatterings is actually extremely small because the photons quickly establish equilibrium with the gas (see also \citealt{furl06-review} for a discussion of the approach to equilibrium).  The heating rate from higher Lyman resonances is also small because those photons are quickly transformed into either \lya photons or continuum photons that easily escape to infinity \citep{pritchard05}.  We therefore neglect this mechanism.  

\subsection{Shock Heating} \label{shockheat}

Finally, we must consider heating from shocks in the IGM.  At low redshifts, shocks play an integral role in its temperature structure \citep{cen99,dave01}.  The shock networks surrounding filaments and halos can be quite complex \citep{miniati00,kang05}, but their characteristic properties are easy to estimate.  The typical shock that has just collapsed at redshift $z$ will have a peculiar velocity equal to the total distance it has collapsed divided by the age of the Universe, $v_{\rm sh} \approx H(z) R_{\rm nl}(z)$, where $R_{\rm nl}$ is the (physical) nonlinear length scale \citep{cen99}.  For a strong shock in a monatomic gas, the corresponding postshock temperature is 
\begin{equation}
T_{\rm sh} = \frac{3 \mu m_p}{16 k_B} \, v_{\rm sh}^2 \approx  \frac{3 \mu m_p}{16 k_B} \, H^2(z) \, R_{\rm nl}^2(z),
\label{eq:shockT}
\end{equation}
where $\mu$ is the mean molecular weight in atomic units.  This yields $T_{\rm sh} \sim 10^7 \kel$ at the present day \citep{cen99}, comparable to the results of numerical simulations.  On the other hand $T_{\rm sh} \sim 10^4 \kel$ at $z \sim 3$, close to the mean temperature of a photoionized medium \citep{furl04-sh}.  Thus large-scale structure shocks decrease in importance at moderate redshifts.

However, \emph{before} reionization they are also important, because without photoheating $T_K$ can be small and the IGM sound speed may be much less than $v_{\rm sh}$.  Unfortunately, studying this regime requires extremely high resolution numerical simulations, and the current generation appears to disagree on their significance.  \citet{gnedin04} found IGM shocks to be the most important factor in determining $\dtb$ at $z \la 15$, but other simulations suggest that filament shocks have a relatively modest effect (comparable to the signal from minihalos; \citealt{kuhlen06-21cm, shapiro05}).  One possible reason for the discrepancy is that the latter two studies assumed that only overdense gas could be observed against the CMB and so only included shocks bounding the cosmic web. \citet{gnedin04}, on the other hand, assumed that the entire IGM was visible because of \lya coupling. Thus it is possible that the tiny sound speed in the IGM allows a population of weak shocks to exist \emph{outside} the cosmic web. Unfortunately, weak shocks in the low density IGM may be difficult to resolve properly in simulations.  Another alternative is that the differences occurred because the studies focused on different redshift regimes.

For now, we will assume the latter and use the simple model of \citet{furl04-sh} to estimate the effect of shocks in the cosmic web (see also \citealt{nath01,valageas02}).  This prescription assumes that shocks form around a region once its mean linearized density exceeds $\delta_{\rm sh}=1.06$.  This is equivalent to assuming that they appear when spherical perturbations ``turnaround" and break off from the Hubble flow (i.e., when the flow begins to converge).  This spherical picture is approximate, but it provides a simple criterion to estimate the fraction of mass that has been shocked on its way to halo formation.  Conveniently, at turnaround, the physical velocity is of course zero, so the peculiar velocity is almost precisely $v_{\rm sh} \approx H(z) R_{\rm nl}(z)$.  Thus it reproduces the characteristic temperature of $z=0$ shocks.  In the same way as the \citet{press74} mass function, it also yields the distribution of shock temperatures at any epoch.  

Thus, by analogy to equation (\ref{eq:fcollps}), the fraction of gas that has been shocked to a temperature $T > T_{\rm sh}$ is
\begin{equation}
f_{\rm sh}(>T_{\rm sh}) = {\rm erfc} \left[ \frac{\delta_{\rm
    sh}(z)}{\sqrt{2} \sigma(T_{\rm sh})} \right],
\label{eq:fsh}
\end{equation}
where $T_{\rm sh}$ is related to the mass of the underlying perturbation via equation (\ref{eq:shockT}).  For example, the mass corresponding to a shock with $T_{\rm sh}=T_\gamma$ is $M_\gamma \approx 4.3 \times 10^5 \Msun$, which is somewhat greater than the minimum collapse mass (the so-called ``filter mass;''  \citealt{gnedin98}) -- though note that $\delta_{\rm sh} < \delta_c$ so more gas will be in the shocked phase than in collapsed objects of a comparable mass.  For our model, we will simply compute the fraction $f_{\rm sh}$ of gas that has been heated to $T_K > T_\gamma$.  We will assume that this gas emits strongly relative to the CMB ($T_S \gg T_\gamma$), while the rest remains at the mean temperature (without shocks).  Note that, once X-rays have heated the IGM above $T_\gamma$, shock heating will become less significant than our prescription assumes; however, in that regime the shocked gas and background IGM emit at nearly the same level and so shocks have almost no observable effect.

We emphasize that this treatment is meant to include only shocks that are part of structure formation (sheets, filaments, and halos).  As described above, there may be a population of weaker shocks in the low-density IGM that deposit a significant amount of thermal energy there.  In the language of our model, these would correspond to shocks that appear \emph{before} turnaround due to random IGM flows.  Our model may thus underestimate $T_K$ at early times and overestimate the amplitude and duration of absorption.  But, as we will see, X-ray heating generally becomes significant at $z \ga 15$, when such perturbations would presumably be significantly smaller than those of \citet{gnedin04}.

\section{The Spin Temperature} \label{tshist}

Once $T_K$ is known, the next step is to compute the spin temperature of the gas.  We include H-H collisions \citep{zygelman05}, H-e$^-$ collisions \citep{smith66, liszt01}, and the Wouthuysen-Field effect \citep{wouthuysen52, field58, hirata05} as described in \S \ref{21cm}.  Collisions are unimportant for the low-density IGM during the epochs of interest,\footnote{Unless, that is, X-ray heating is extremely strong and renders $\bxion \ga 0.03$ \citep{nusser05-xray}.}, so the Wouthuysen-Field mechanism is therefore the most important factor.  Its strength depends on the soft-UV radiation background, for which we will construct a simple, easily parameterized model.  We again assume that young stars dominate the flux, so that we can write the comoving emissivity at frequency $\nu$ is
\begin{equation}
\epsilon(\nu,z) = f_\star \, \bar{n}_b^0 \, \epsilon_b(\nu) \,
\frac{\deriv \fcoll}{\deriv t}.
\label{eq:sfemiss}
\end{equation}
The function $\epsilon_b(\nu)$ is the number of photons produced in the frequency interval $\nu \pm \deriv \nu/2$ per baryon incorporated into stars.  For spin temperature coupling, we are interested only in photons between \lya and the Lyman-limit.\footnote{Here we ignore Lyman-line photons produced inside galaxies as well as line photons produced following IGM ionizations.  The former is a reasonable assumptions because these photons rapidly redshift out of resonance and hence only affect a small fraction of the IGM (much of which is predominantly ionized).  The second is an excellent approximation for Pop II stars but less good for Pop III stars, because of their hard spectra.  Nevertheless, such photons also rapidly redshift out of resonance near the edge of the \htwo regions.}  We use the spectral models of \citet{barkana05-ts}, which fit a power law between each Lyman-series resonance.  For Population III stars, a total of $N_\alpha=4800$ photons are produced per baryon in the interval $(\nu_\alpha,\nu_{\rm LL})$, with a relatively flat spectral distribution \citep{bromm01-vms}.  For continuously forming low-metallicity Population II stars, $N_\alpha=9690$ photons are produced per baryon, with the spectrum falling steeply past the Ly$\beta$ resonance \citep{leitherer99}. 

The average \lya background is then
\begin{eqnarray}
J_\alpha(z) & = & \sum_{n=2}^{n_{\rm max}} J_\alpha^{(n)}(z) \nonumber \\
& = & \sum_{n=2}^{n_{\rm max}} f_{\rm rec}(n) \int_z^{z_n'} 
\deriv z' \frac{(1+z)^2}{4 \pi} \frac{c}{H(z')} \epsilon(\nu_n',z'),
\label{eq:jalpha}
\end{eqnarray}
where $J_\alpha^{(n)}$ is the background from photons that originally redshift into the \lyn resonance, $\nu_n'$ is the frequency at redshift $z'$ that redshifts into that resonance at redshift $z$,
\begin{equation}
\nu_n' = \nu_n \frac{(1+z')}{(1+z)},
\label{eq:nun}
\end{equation}
and
\begin{equation}
\frac{1 + z_n'}{1+z} = \frac{1 - (n+1)^{-2}}{1-n^{-2}}
\label{eq:zmax}
\end{equation}
is the largest redshift from which a photon can redshift into it.  The factor $f_{\rm rec}(n)$ is the fraction of the resulting \lyn photons that actually cascade through \lya and induce strong coupling \citep{hirata05, pritchard05}.  We truncate the sum at $n_{\rm max}=23$ (following \citealt{barkana05-ts}), although the closely-spaced $n \gg 1$ levels contribute only a small fraction of the total flux.  Finally, we compute the coupling coefficient $\tilde{x}_\alpha$ from equation (\ref{eq:xalpha}).  

For estimation purposes, we can approximate equation (\ref{eq:jalpha}) as
\begin{equation}
J_\alpha \approx \frac{c}{4 \pi} \, \bar{f}_{\rm rec} \, f_\star \,
\bar{n}_b^0  \, \Delta \fcoll \, \frac{N_\alpha}{\Delta \nu} \,
(1+z)^2. 
\label{eq:jalpha-approx}
\end{equation}
Here $\bar{f}_{\rm rec}$ is the average probability that a photon in the interval $(\nu_\alpha,\nu_{\rm LL})$ is converted into a \lya photon.  It is $\bar{f}_{\rm rec}=0.63$ and $0.72$ for our Pop III and Pop II spectra \citep{pritchard05}.  We have approximated the spectrum as constant (so that $\epsilon_b \sim N_\alpha/\Delta \nu$, where $\Delta \nu =\nu_{\rm LL} - \nu_\alpha$) and let $\Delta \fcoll
\sim \fcoll$ be the spectrum-weighted collapse fraction over the appropriate redshift intervals in equation (\ref{eq:jalpha}).  

\section{The Ionization History} \label{ionhist}

The next step in understanding the global history of the 21 cm signal is $\bxion(z)$.  This has been a subject of extensive study over the past three decades (see, e.g., \citealt{barkana01,haiman04-rev,ciardi05-rev} for recent reviews).  As such, we will not attempt to explore all of its aspects.  We will again examine a simple model encapsulating the major features in the global evolution.

We associate reionization with the star formation rate in a similar way to the \lya and X-ray backgrounds (see eqs. [\ref{eq:xrayemiss}] and [\ref{eq:sfemiss}]).  We write 
\begin{equation}
\frac{\deriv \bxion}{\deriv t} = \zeta(z) \frac{\deriv \fcoll}{\deriv t}
- \alpha_A C(z,\xion) \bxion(z) \bar{n}_e(z),
\label{eq:bxion-evol}
\end{equation}
where $\zeta$ is an ionizing efficiency parameter, $\alpha_A = 4.2 \times 10^{-13} \recunits$ is the recombination coefficient at $T=10^4 \kel$,\footnote{Note that we use the case-A value, which essentially amounts to assuming that ionizing photons are absorbed inside dense, neutral systems \citep{miralda00}.}  $C \equiv \VEV{n_e^2}/\VEV{n_e}^2$ is the clumping factor, $\bar{n}_e$ is the average electron density in ionized regions, and $\zeta$ is an ionizing efficiency parameter.  The first term on the right hand side is the source term for ionizing photons and the second term is a sink describing recombinations. 

We must now specify $\zeta$, $\mmin$ (for $\fcoll$), and $C$.  The ionizing efficiency can be written
\begin{equation}
\zeta = A_{\rm He} f_\star f_{\rm esc} N_{\rm ion},
\label{eq:zetadefn}
\end{equation}
where $f_{\rm esc}$ is the fraction of ionizing photons that escape their host galaxy into the IGM, $N_{\rm ion}$ is the number of ionizing photons per baryon produced in stars, and $A_{\rm He}$ is a correction factor for helium.  With the exception of $A_{\rm He}$, all of these parameters are highly uncertain.  The star formation efficiency depends on the dynamics of star formation in young galaxies, $f_{\rm esc}$ depends on the distribution of gas, dust, and stars within galaxies, and $N_{\rm ion}$ depends on the stellar IMF and metallicity.  We will therefore content ourselves with representative values of these parameters in our calculations (see the discussion in \citealt{furl06-review} for more information).

Of course, both $\zeta$ and $\mmin$ could evolve with time because of feedback.  For example, metal enrichment could decrease $\zeta$ if Pop II stars are less efficient than their Pop III progenitors, or $\mmin$ could increase because of photoheating during reionization (\citealt{rees86, efstathiou92, thoul96,kitayama00}; but see \citealt{dijkstra04}).  Feedback is one promising method to slow reionization and reconcile the SDSS quasar data with \emph{WMAP}, but the mechanisms at play and their interactions are essentially unconstrained.  For our purposes, the important question is how they affect the global reionization history.  Early efforts suggested that they might introduce sharp features such as ``double reionization" \citep{cen03, wyithe03}.  However, this requires a sharp drop in the ionizing efficiency that is synchronized across the entire Universe.  That is difficult to arrange when the inhomogeneities inherent to local feedback mechanisms are included \citep{furl05-double}.  Instead, feedback is likely to prolong reionization or introduce a ``stalling" phase.  We will show some examples generated using the feedback models of \citet{furl05-double} to illustrate its effects. 

The clumping factor can, in principle, be computed through numerical simulations with radiative transfer.  Unfortunately, that requires fully resolving small-scale structure in the IGM as well as its evolution throughout reionization.  Moreover, $C$ depends on how the IGM is ionized -- for instance, whether low-density gas is ionized first or whether many photons are consumed ionizing dense blobs when $\bxion$ is still small \citep{furl05-rec}.  Thus $C$ must be recalculated for each different set of model parameters.  Unfortunately, simulations have not yet achieved these goals.  Following our usual policy, we will therefore use a simple analytic model \citep{miralda00}; in this prescription, voids are ionized first.  This probably underestimates the mean recombination rate, so reionization may be a bit slower than our models predict (and begin a bit later).  On the other hand, simulations suggest that recombinations do not dramatically slow reionization, even if minihalos are included \citep{ciardi05-mh}.  Note that when recombinations are slow, equation (\ref{eq:bxion-evol}) can be approximated as
\begin{equation}
\bxion \approx \frac{\zeta \fcoll}{1 + \bar{n}_{\rm rec}},
\label{eq:bxion}
\end{equation}
where $\bar{n}_{\rm rec}$ is the mean cumulative number of recombinations per hydrogen atom.

\section{Critical Points in the 21 cm History}
  \label{critpt}

Now that we have assembled the relevant physics, we are ready to construct some plausible histories.  But first it is useful to identify three ``critical points" in the 21 cm history: the redshift $z_h$ at which the IGM is heated above $T_\gamma$, the redshift $z_c$ at which the Wouthuysen-Field mechanism couples first $T_S$ and $T_K$ (when $\tilde{x}_\alpha=1$), and the redshift of reionization $z_r$.  Their relative timing determines the structure of the 21 cm spectrum.  

The most important question is whether $z_c$ occurs before the other two transitions:  if not, the IGM will remain invisible.  We can compute the collapse fraction (or more precisely $\Delta \fcoll$) required to achieve $\tilde{x}_\alpha=1$ from equation (\ref{eq:jalpha-approx}).  Substituting this into equation (\ref{eq:xrayemiss}), integrated over time, we have the net X-ray heat input $\Delta T_c$ at $z_c$:
\begin{eqnarray}
\frac{\Delta T_c}{T_\gamma} & \sim 0.08 & f_X \left( \frac{f_{X,h}}{0.2} \,
\frac{\fcoll}{\Delta \fcoll} \, \frac{9690}{N_\alpha} \,
\frac{1}{\tilde{S}_\alpha} \, \frac{0.6}{\bar{f}_{\rm rec}} \,
  \frac{0.023}{\Omega_b h^2} \right) \nonumber \\ 
& & \times \left( \frac{20}{1+z} \right)^3.
\label{eq:dtc}
\end{eqnarray}
Note that $\Delta T_c$ is independent of $f_\star$ because both the coupling and heating rates are proportional to the star formation rate (by assumption, of course).  Interestingly, for fiducial (Pop II) parameters $z_c$ should occur somewhat before $z_h$ (see also the simple models of \citealt{chen04,hirata05}).  For Pop III stars, $N_\alpha$ is a factor of two smaller and $f_X$ may be significantly larger than unity, in which case absorption would be much less important.  Thus the presence or absence of a strong and extended absorption epoch offers a useful probe of high-redshift star formation.

A similar estimate of the ionization fraction at $z_c$ yields
\begin{eqnarray}
\bar{x}_{i,c} & \sim 0.05 & \left( \frac{\fesc}{1+\bar{n}_{\rm rec}} \,
\frac{N_{\rm ion}}{N_\alpha} \, \frac{\fcoll}{\Delta \fcoll} \,
\frac{1}{\tilde{S}_\alpha} \, \frac{0.6}{\bar{f}_{\rm rec}} \,
\frac{0.023}{\Omega_b h^2} \right) \nonumber \\ 
& & \times \left( \frac{20}{1+z} \right)^2.
\label{eq:xic}
\end{eqnarray}
For Pop II stars, $N_{\rm ion}/N_\alpha \approx 0.45$; thus even in the worst case ($\fesc=1$ and $\bar{n}_{\rm rec}=0$) coupling would become efficient during the initial stages of reionization.  However, Pop III stars have much harder spectra and $N_{\rm ion}/N_\alpha \approx 7$.\footnote{Note that Fig. 1 of \cite{ciardi03-sim} seems to suggest much smaller ratios for Pop III stars; this is because they report the number of photons \emph{per frequency interval}.  The frequency range between \lya and the Lyman limit is much smaller than that available to ionizing photons, which accounts for the difference.}  In principle, it is therefore possible for Pop III stars to reionize the universe \emph{before} $z_c$.  This requires a small clumping factor, relatively late reionization ($z \la 10$), and $\fesc \approx 1$.  There is some theoretical motivation for such a large escape fraction in the earliest generation of galaxies \citep{whalen04}, but maintaining that in larger galaxies seems unlikely (as might widespread and dominant Pop III star formation at $z<10$).  More conventional histories, in which $\fesc \la 0.2$ or $z_r \ga 12$ for Pop III stars, predict strong coupling throughout reionization (cf. \citealt{ciardi03-21cm}).

Finally, it is interesting to check whether we will see the IGM in absorption or emission during reionization.  Combining equations (\ref{eq:xrayemiss}) and (\ref{eq:bxion}), we find that the total heat input as a function of $\bxion$ is
\begin{equation}
\frac{\Delta T(\bxion)}{T_\gamma} \sim \left( \frac{\bxion}{0.025} \right) \,
\left( f_X \, \frac{f_{X,h}}{\fesc} \, \frac{4800}{N_{\rm ion}} \,
\frac{10}{1+z} \right) \, (1 + \bar{n}_{\rm rec}).
\label{eq:tx-xi}
\end{equation}
Thus, provided $f_X \ga 1$ and Pop II stars dominate, it is safe to assume that the IGM is much warmer than the CMB throughout reionization.  However, because Pop III stars have much higher ionizing efficiencies, it is possible for X-ray heating to become significant only when $\bxion \ga 0.2$; in this case we would be able to see the early stages in absorption so long as $f_X$ remains small.  This would be exciting in that it would add extra structure to the reionization signal, but it would also make isolating $\bxion(z)$ more difficult.

%%%%%%%%%%%% FIGURE 1: Global history, pop II 
\begin{figure*}[!t]
\begin{center}
\resizebox{8cm}{!}{\includegraphics{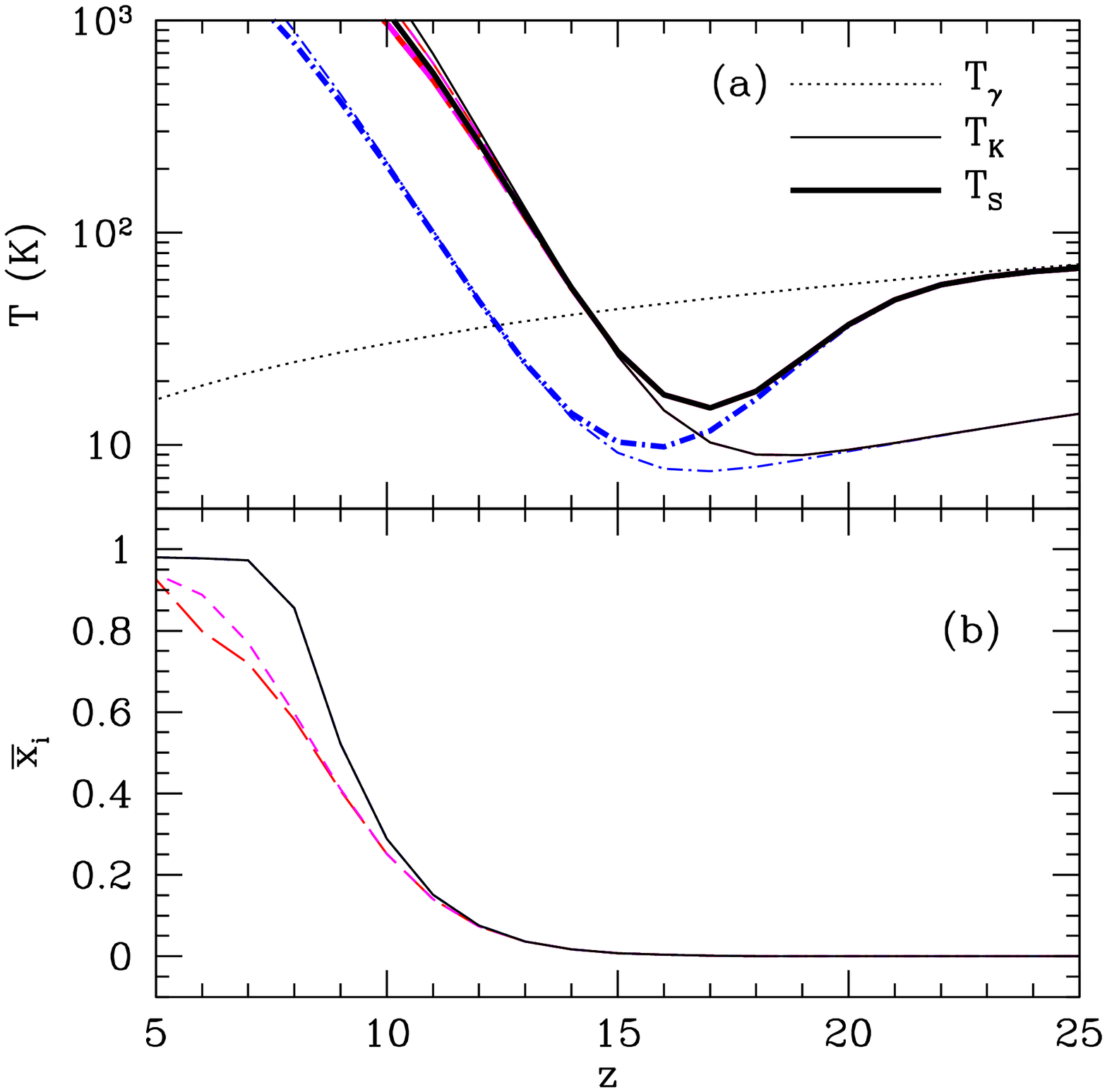}}
\hspace{0.13cm}
\resizebox{8cm}{!}{\includegraphics{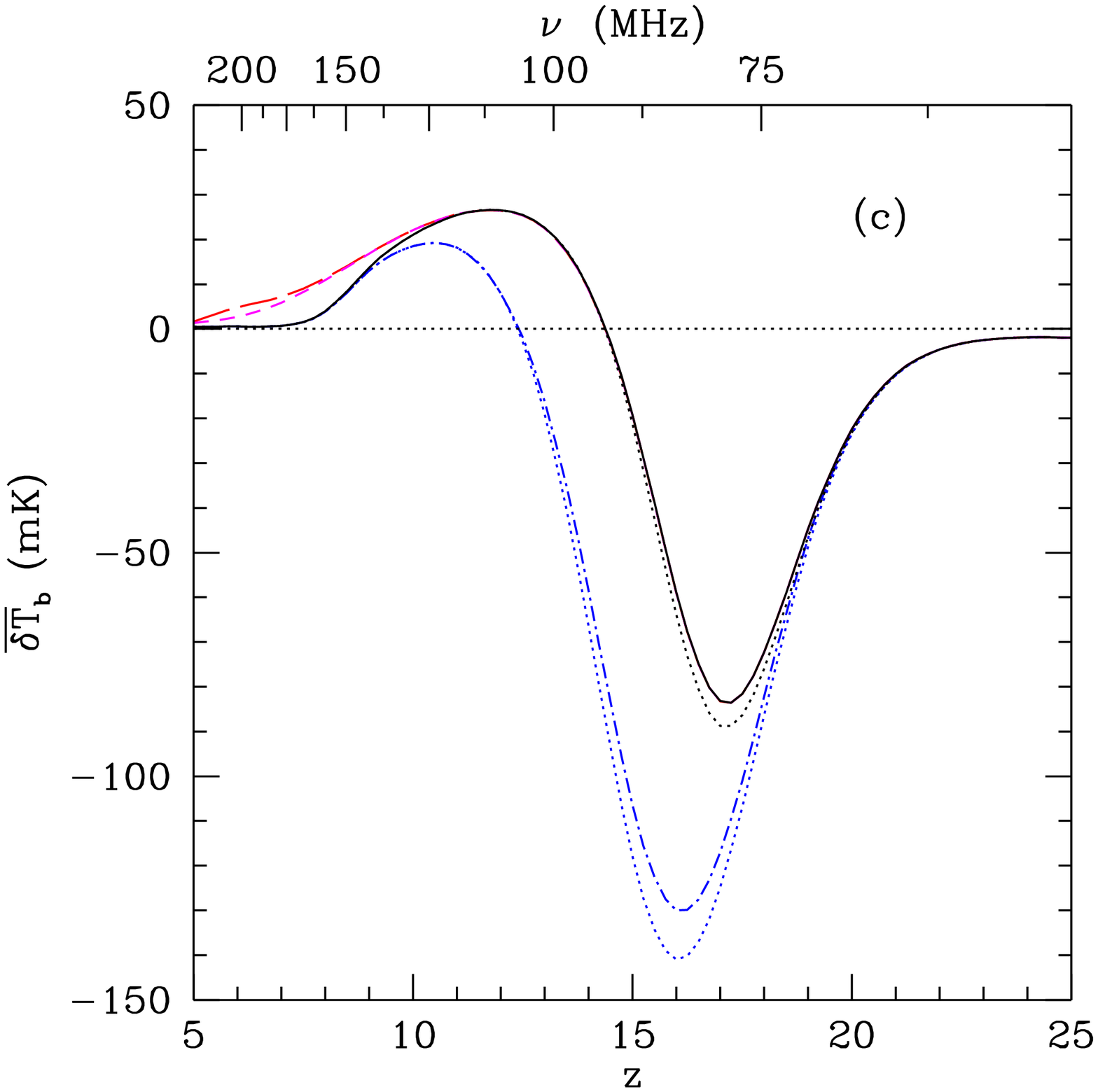}}
%\plottwo{f3-2ab.eps}{f3-2c.eps}
\end{center}
\caption{Global IGM histories for Pop II stars.  The solid curves take our fiducial parameters.  The dot-dashed curves take $f_X=0.2$.  The short- and long-dashed curves include photoheating feedback (see text). \emph{(a)}: CMB, gas kinetic, and spin temperatures (dotted, thin, and thick curves, respectively). \emph{(b)}: Ionized fraction.  \emph{(c)}: Differential 21 cm brightness temperature against the CMB.  In this panel, the two dotted lines show $\delta T_b$ if shock heating is ignored.}
\label{fig:pop2-glob}
\end{figure*}

\section{Global Histories} \label{glob-theor}

We will now construct some example histories using the prescriptions given in \S \ref{therm}--\ref{ionhist}.  Of course, given the uncertainties in each of the parameters, making firm predictions is all but impossible.  But we can highlight the possible range of features.  We will necessarily neglect a number of potentially important effects (such as metal enrichment and the evolution of the H$_2$ cooling efficiency), but we hope that our histories are nevertheless representative of the possibilities.

\subsection{Pop II Stars} \label{pop2}

We begin with a fiducial set of Pop II parameters.  We take $\mmin=m_4$, $f_\star=0.1$, $\fesc=0.1$, $f_X=1$, $N_{\rm ion}=4000$, and $N_\alpha=9690$.  (Thus $\zeta=40$ for this model.)  The solid curves in Figure~\ref{fig:pop2-glob} show the resulting global history.  In panel \emph{(a)}, the dotted curve is $T_\gamma$, the thin curves are $T_K$, and the thick curves are $T_S$ (note that they always interpolate between $T_\gamma$ and $T_K$).  

Figure~\ref{fig:pop2-glob}{\em b} shows the ionization history.  Without feedback, it increases smoothly and rapidly over a redshift interval of $\Delta z \sim 5$.  For these parameters, reionization is essentially complete at $z_r \sim 7$.  Other values of $\zeta$ move $z_r$ back and forth but do not strongly affect its width in redshift space.  Changing $f_\star$ affects all three processes (nearly) equally:  it simply moves the entire curve forward or backward in redshift.  Note that the evolution slows somewhat at $\bxion \ga 0.8$ and never quite reaches complete reionization; this is because the clumping factor $C$ increases rapidly as $\bxion \rightarrow 1$ \citep{miralda00}.  The dashed curves in this panel use photoheating feedback to slow reionization.  Each assumes that the minimum virial temperature increases to $T_h=2 \times 10^5 \kel$ in ionized regions, near the upper limit of theoretical expectations \citep{dijkstra04-feed}.  All other parameters are assumed to remain constant in photoheated regions.  The two curves choose different prescriptions for the effective volume in which the feedback occurs; they bracket the possible effects of photoheating \citep{furl05-double}.\footnote{In detail, the short-dashed curve assumes that the photoheated volume is identical to the ionized volume, while the long-dashed curve calculates the photoheated volume without including recombinations.  See \citet{furl05-double}.}  Clearly, $\bxion$ evolves monotonically even with this relatively powerful negative feedback.  It does, however, slow the pace of reionization when $\bxion \ga 0.5$, spreading it over a significantly longer time interval.\footnote{In fact, with these parameters feedback pushes $z_r \la 5$, which is ruled out observationally.  But that is of no consequence to us because we only wish to examine its qualitative effects on the 21 cm signal.}

Figure~\ref{fig:pop2-glob}{\em c} shows the 21 cm brightness temperature increment $\dtb$ relative to the CMB.  Here for convenience we have also labeled the corresponding observed frequency $\nu=\nu_{21}/(1+z)$, where $\nu_{21}=1420 \MHz$.  The signal clearly has interesting structure.  The most robust feature is the $\sim 30 \mkel$ drop in $\dtb$ during reionization.  For our fiducial parameters (and without feedback), that spans a frequency interval of $\sim 35 \MHz$.  In this model, recombinations are slow (delaying reionization by only $\Delta z \approx 1$) and $\bxion$ is essentially proportional to $\fcoll$.  Thus, unless the ionizing efficiency {\em increases} throughout reionization because of some positive feedback mechanism (or if more massive galaxies drive reionization; \citealt{furl05-charsize}), the signal is unlikely to have a much larger gradient during reionization.  But with feedback, it can be significantly smaller: in that case, $\dtb$ falls to zero over $\Delta z \approx 8$ ($\sim 60 \MHz$ here).  It is this gradient that might allow us to isolate the cosmological signal, so feedback will make its detection rather more difficult.

%%%%%%%%%%%% FIGURE 2: Global history, pop III
\begin{figure*}[!t]
\begin{center}
\resizebox{8cm}{!}{\includegraphics{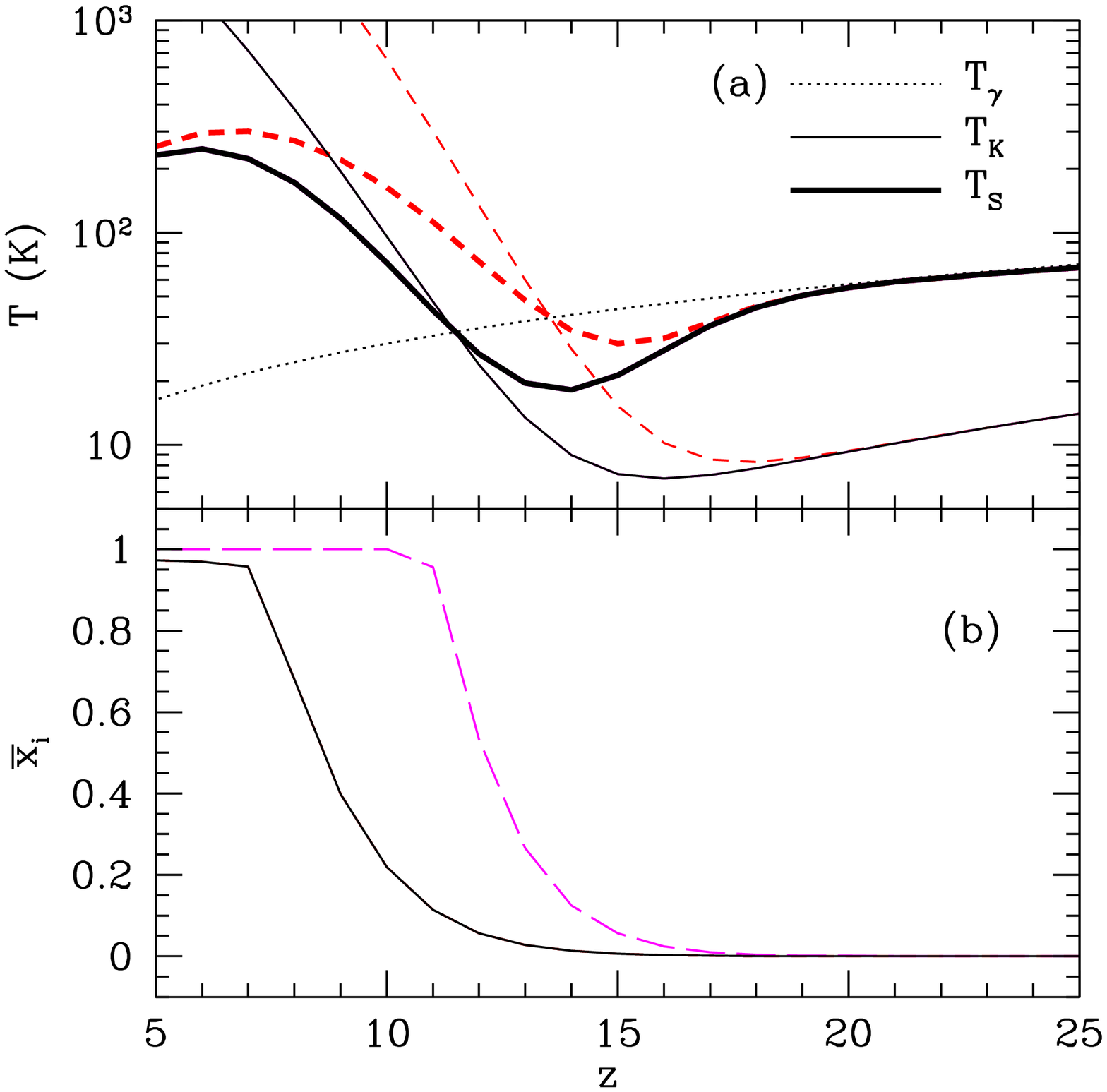}}
\hspace{0.13cm}
\resizebox{8cm}{!}{\includegraphics{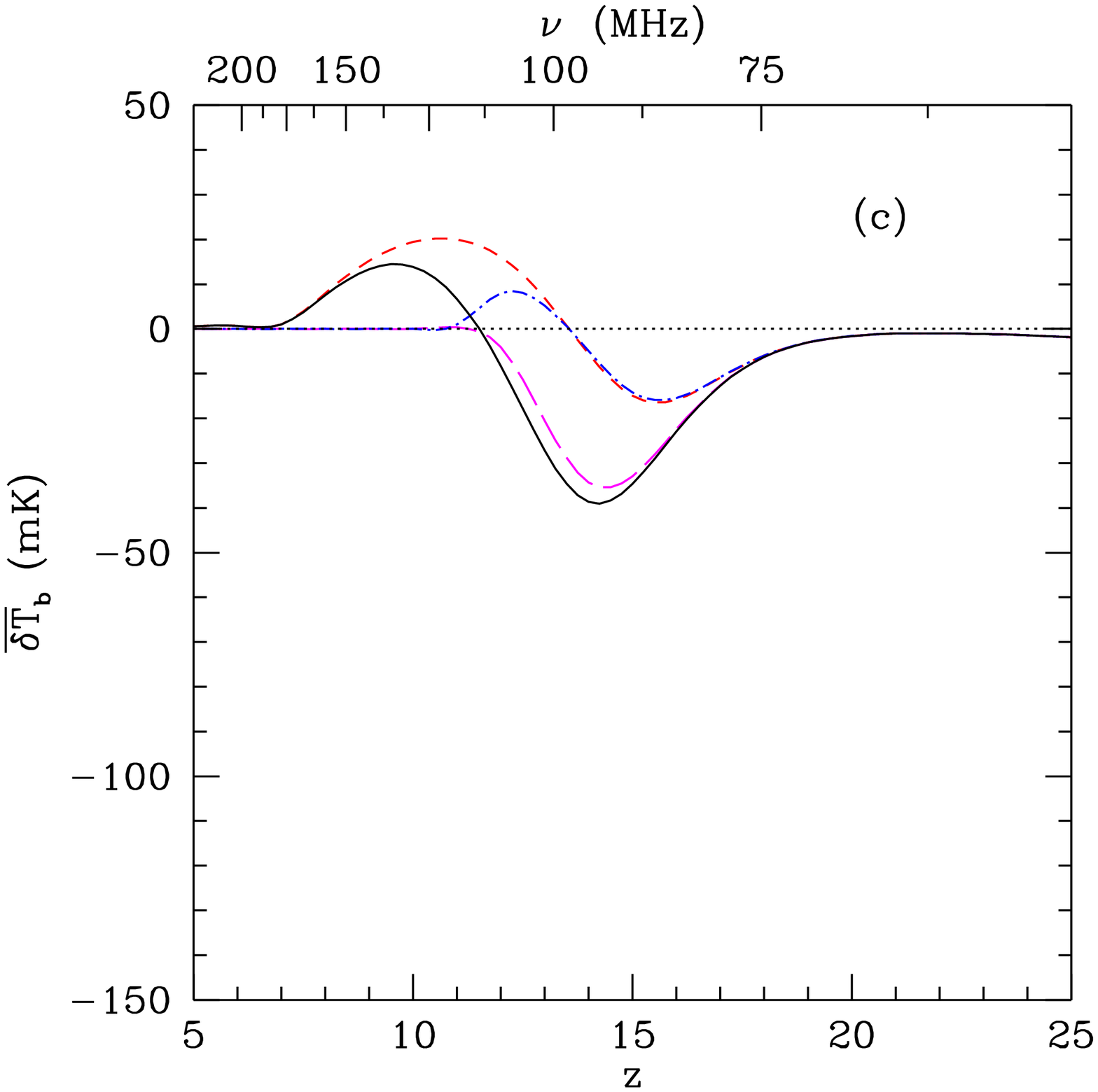}}
%\plottwo{f3-2ab.eps}{f3-2c.eps}
\end{center}
\caption{Global IGM histories for very massive Pop III stars.  Panels are the same as in Fig.~\ref{fig:pop2-glob}.  The solid curve takes our fiducial Pop III parameters (without feedback).  The long-dashed lines take $\fesc=1$, the short-dashed lines take $f_X=5$, and the dot-dashed line (shown only in {\em c}) assumes $\fesc=1$ and $f_X=5$.}
\label{fig:pop3-glob}
\end{figure*}

A stronger feature appears near $z_c$ and $z_h$.  At $z \ga 20$, the IGM is nearly invisible even though $T_K \ll T_\gamma$.  Once the first galaxies form, the Wouthuysen-Field effect becomes important and $T_S$ drops rapidly.  Because $N_\alpha$ is large for Pop II stars, efficient coupling precedes the heating transition, allowing a relatively strong absorption signal ($\dtb \approx -80 \mkel$) to appear by $z \sim 20$. Absorption persists until $z \sim 15$ (or $\nu \sim 90 \MHz$) before fading
into emission.  Thus, for these fiducial parameters, the universe is cold when coupling becomes strong, but it still heats up well before reionization begins in earnest.  The absorption epoch can therefore teach us about the appearance and spectral properties of the first sources of light in the universe \citep{tozzi00,sethi05}, especially the relative importance of Wouthuysen-Field coupling and X-ray heating.  This era also produces much sharper features than reionization itself:  the signal changes by $\sim 110 \mkel$ over $\Delta z \approx 5$ (or $\Delta \nu \approx 20 \MHz$ at such high redshifts).  Thus, even though the foregrounds are also much larger at the lower frequencies, the early evolution may not be much harder to identify (especially if feedback substantially slows reionization).

Note that our model predicts a significantly stronger absorption epoch than \cite{chen04}; the difference lies in the assumed $f_X$ and in their assumption of a \emph{linearly} increasing emissivity throughout reionization.  In reality, structure formation proceeds exponentially at high redshifts, so the X-ray (and ionizing) luminosities most likely decline much more rapidly with redshift than assumed in their model.  This permits a longer epoch of absorption.

The other curves in Figure~\ref{fig:pop2-glob} illustrate how the evolution changes with some of our input assumptions.  First, the dot-dashed curve assumes $f_X=0.2$ (near the value of \citealt{glover03}).  This has no effect on $\bxion$, but it delays the heating transition by $\Delta z \sim 3$.  As a result, $T_S$ remains relatively close to $T_\gamma$ during the beginning of reionization, and $\dtb$ never reaches its peak value in emission.  On the other hand, the absorption trough is deeper, though it is still confined to $z \ga 15$.  Also, we emphasize that the IGM remains visible throughout reionization, even when $T_S \la T_\gamma$.

The two dotted curves in panel {\em (c)} show $\dtb$ if we ignore shock heating.  In this case, all of the IGM material remains cold.  At the high redshifts ($z \ga 15$) that are relevant here, shock heating in filaments affects $\la 10\%$ of the IGM, so its effects on $\dtb$ are modest.  If weak shocks do occur outside of collapsing structures, the absorption epoch will be much weaker.  Moreover, if heating is delayed until lower redshifts where structure formation has progressed much further, shocks should have a much larger effect \citep{gnedin04,furl04-sh}.

%%%%%%%%%%%%% FIGURE 3: Global history, ``double reionization''
\begin{figure*}[!t]
\begin{center}
\resizebox{8cm}{!}{\includegraphics{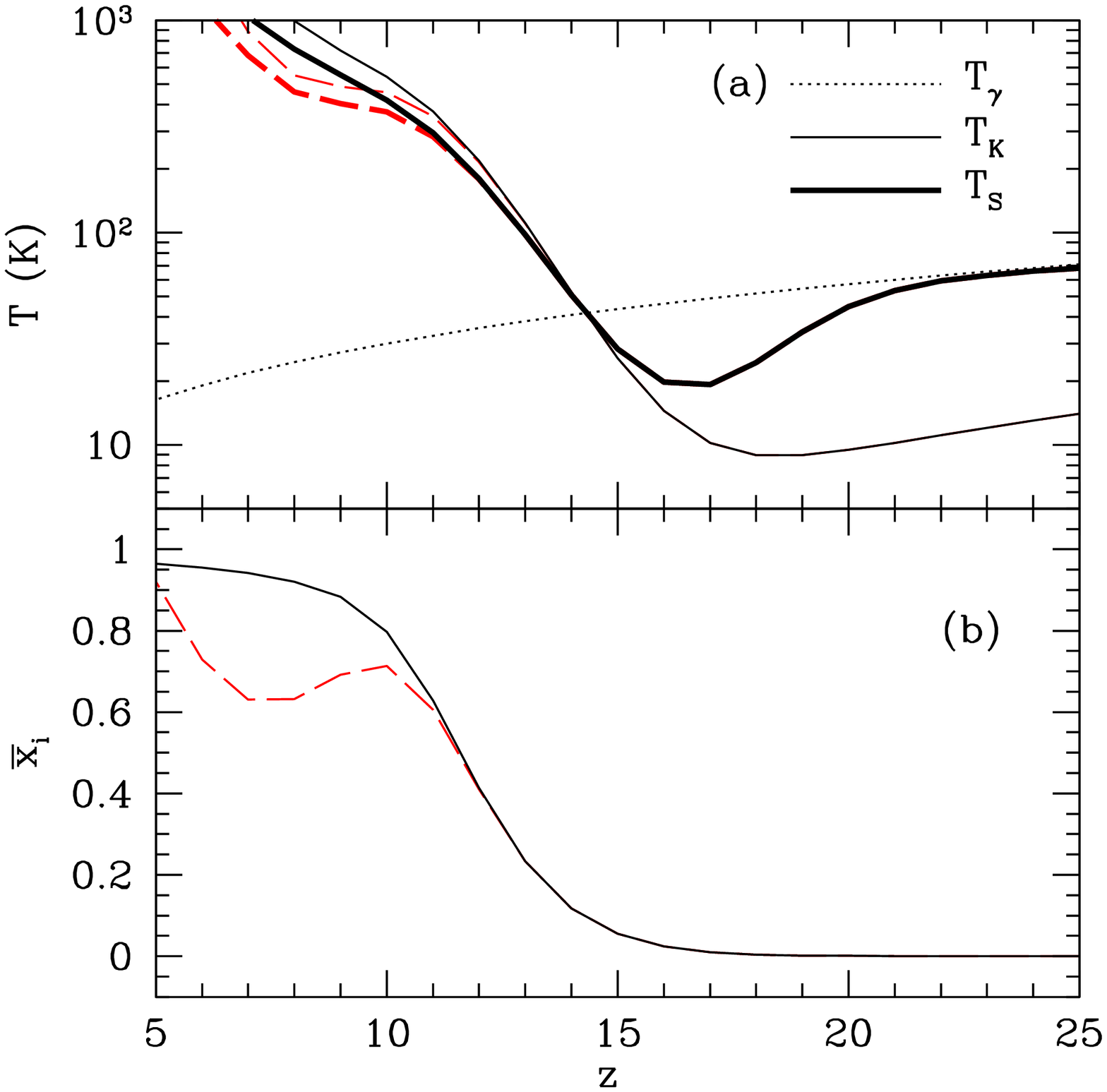}}
\hspace{0.13cm}
\resizebox{8cm}{!}{\includegraphics{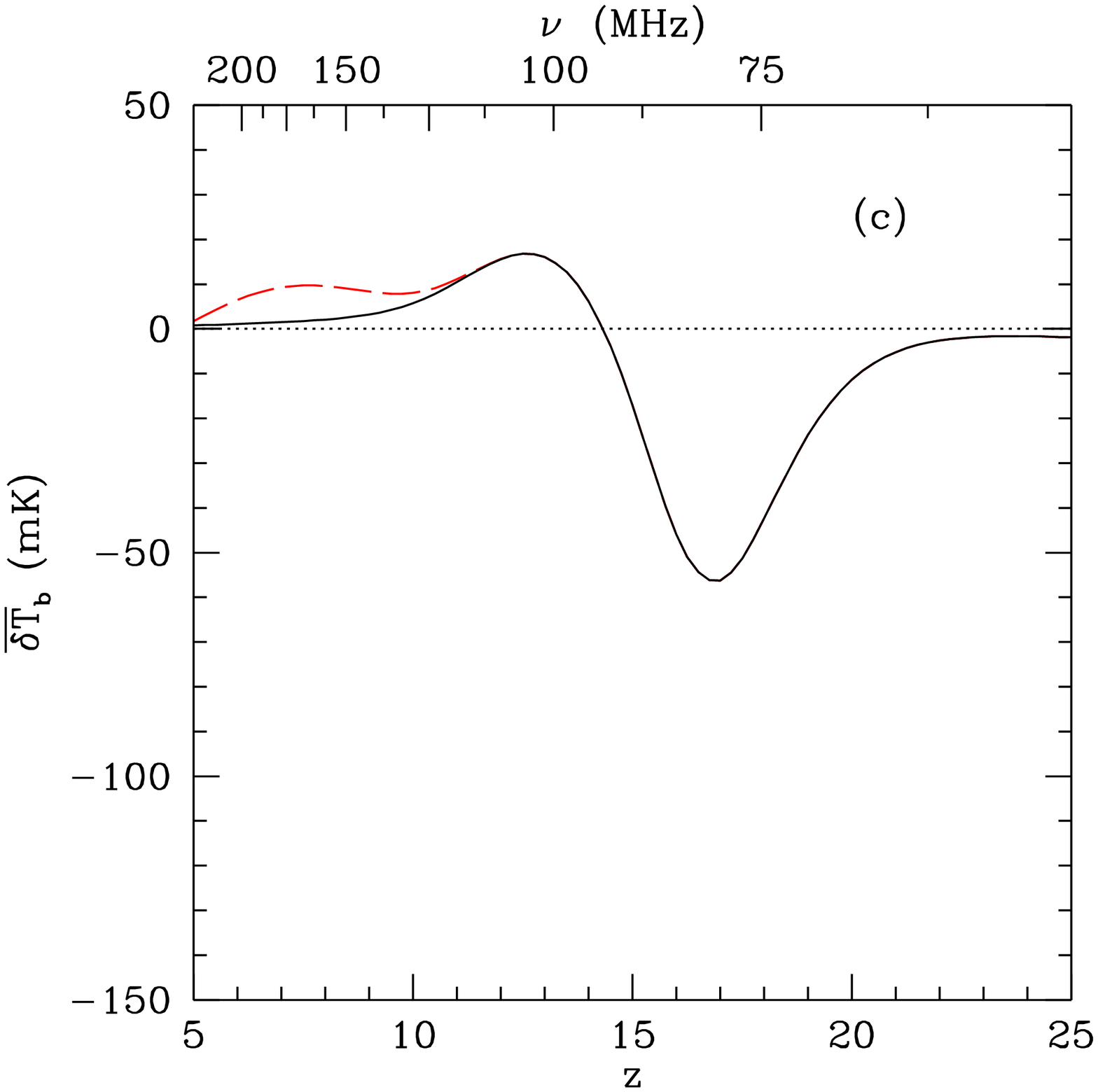}}
%\plottwo{f3-2ab.eps}{f3-2c.eps}
\end{center}
\caption{Global IGM histories for a two-population model.  Panels are the same as in Fig.~\ref{fig:pop2-glob}.  Both sets of curves assume Pop III stars form in cold regions and Pop II stars form in  hot regions (each with our fiducial parameters, except that $f_\star=0.1$ for Pop III stars).}
\label{fig:poptrans-glob}
\end{figure*}

\subsection{Population III Stars} \label{pop3}

Figure~\ref{fig:pop3-glob} shows similar histories for Pop III stars.  Here we take fiducial parameters (shown by the solid curves) $\mmin=m_4$, $f_\star=0.01$, $\fesc=0.1$, $f_X=1$, and $N_{\rm ion}=30,000$, yielding $\zeta=30$; reionization ends at $z \approx 7$ in this case, ignoring feedback.  The thermal history is similar to the Pop II case, except that the coupling and heating transitions are much more closely spaced in time because Pop III stars have relatively few soft-UV photons per high energy photon.  As a result, $T_S \rightarrow T_K$ occurs just about when $T_K \sim T_\gamma$.  Thus, if Pop III stars dominate, the absorption epoch will likely be considerably weaker -- with a peak signal $|\dtb| \la 50 \mkel$.  In general, with Pop III stars, we find that the time interval over which each quantity evolves is similar to the Pop II case, but the peak signals are up to a factor of two weaker because their spectra push the three transition points closer together.  Thus detecting the global signal will be somewhat harder if massive Pop III stars dominate throughout reionization.  

The other curves use different input parameters.  The long-dashed curves take $\fesc=1$ (which may be appropriate for the first stars, \citealt{whalen04}, although it is likely an overestimate in larger galaxies).  In this case the universe is obviously ionized much earlier, but the thermal history remains the same.  As a result, reionization occurs while the IGM is still cold (see \S \ref{critpt}).  This would complicate the interpretation of $\dtb(z)$ (because temperature fluctuations could be significant) but actually makes the reionization signal slightly stronger.  The more important point is that the Wouthuysen-Field effect is already strong, so fluctuations will appear throughout reionization:  even though $N_{\rm ion}/N_\alpha \approx 7$ for these stars, coupling precedes reionization because each \lya photon is scattered many times.  As suggested by equation (\ref{eq:xic}), arranging it so that $z_c < z_r$ requires extremely unusual parameters, so it is almost impossible to imagine that the IGM would be invisible during reionization.

The short-dashed curves assume $f_X=5$.  This strongly increases the heating rate, further decreasing the strength and duration of the absorption epoch.  On the other hand, this choice cleanly separates the reionization and absorption epochs and yields a brief period where $\bxion$ is small but the IGM strongly emits.  It gives a similar total heating rate to \cite{hirata05}, who also predict weak absorption.

Finally, the dot-dashed curve (shown only in panel {\em c}) takes $\fesc=1$ and $f_X=5$.  This is, in many ways, a ``worst case scenario'' for studying reionization, because it compresses all of our critical points into a short time interval.  Nevertheless, there remains interesting structure during and even before reionization.  The difference is that separating the effects of $T_S$, $T_K$, and $\bxion$ will be more difficult.  

\subsection{A Dual Population History} \label{twopop}

In Figure~\ref{fig:poptrans-glob}, we exaggerate the strength of the photoheating feedback transition by assuming that Pop III stars form in ``cold" regions while Pop II stars form in hot regions.  Both populations take our fiducial parameters except that we set $f_\star=0.1$ for Pop III stars.  Thus the ionizing efficiency declines from $\zeta^{\rm III}=300$ to $\zeta^{\rm II}=40$ during photoheating, and $\mmin$ simultaneously increases.  There is no real physical justification for conflating the metallicity transition with photoheating -- in fact it is much  more gradual than photoheating and requires fine-tinuing to synchronize enrichment and reionization \citep{furl05-double} -- but it provides a simple way to implement exceptionally strong feedback.  As in Figure~\ref{fig:pop2-glob}, the solid and dashed curves bracket the effectiveness of photoheating feedback.  In the fast case, there is a short period in which $\bxion$ overshoots because of the dramatic decline in ionizing efficiency between the Pop III and Pop II stars.  The resulting recombination epoch is nevertheless shallow, and producing true ``double reionization" scenarios requires pushing parameters in even more uncomfortable directions.

Nevertheless, Figure~\ref{fig:poptrans-glob} shows that strong feedback can dramatically slow the later stages of reionization.  For obvious reasons, photoheating feedback becomes important when $\bxion \sim 0.5$.  Thus the gradient remains more or less unchanged during the first half of reionization (and the absorption epoch is completely unaffected), but it decreases strongly in the middle stages.  Any potential ``recombination epoch" is also difficult to see, because $\bxion$ changes by only $\sim 15\%$ over $\Delta z \sim 4$ (yielding $d \dtb/d \nu \sim 0.1 \mkel \MHz^{-1}$).  On the other hand, if feedback is fast, the late stages of reionization are completed entirely by the Pop II objects and so have as large a gradient as the single-population models.  Thus strong feedback could perhaps be inferred by observing two widely separated periods with large gradients.

\section{Discussion} \label{disc}

In this paper, we have described the basic mechanisms relevant to computing the globally-averaged 21 cm signal from high redshifts.  The basic structure of the background depends on three crucial transition points: when $T_S \rightarrow T_K$ (due to the Wouthuysen-Field effect), when $T_K$ first exceeds $T_\gamma$ (most likely due to X-ray heating, possibly together with weak shocks), and reionization.  The relative timing of these three transitions determines how much we can learn from the 21 cm transition.

We have examined the range of possible signals and emphasized their qualitative features.  If Pop II stars dominate the background, and if their X-ray emission (per unit star formation rate) is comparable to that in the local Universe ($f_X \approx 1$), \lya coupling will become efficient before X-ray heating.  This leads to a relatively strong absorption epoch (with a peak $|\dtb| \sim 100 \mkel$) until X-rays cause $T_K \gg T_\gamma$.  Once that occurs, $\dtb$ saturates until reionization begins -- which is well after the heating transition so long as $f_X \ga 1$.  Such histories contain long periods in which density fluctuations dominate the 21 cm signal, which is useful for studying cosmological parameters \citep{bowman05-param,mcquinn05-param}.  This points to an important benefit of our calculation:  it provides a basis for evaluating the expected amplitude of the 21 cm \emph{fluctuations} during various phases of structure formation.  Cosmological measurements, spin temperature fluctuations \citep{barkana05-ts, pritchard05}, minihalos \citep{iliev02, furl06-mh}, and reionization itself must all be placed in their proper contexts, and this is a first step in that direction.

If very massive Pop III stars dominate reionization, we expect a rather different history.\footnote{If metal-free stars have a more or less normal Salpeter IMF, they produce histories resembling those of our Pop II models.}  Because these stars are so hot, the number of \lya photons per ionizing photon is much smaller.  Thus the \lya coupling transition occurs much nearer to reionization, although the reionization epoch will still be visible in the 21 cm transition.  With these stellar populations, disentangling the effects of $T_S$ and $\bxion$ may be more difficult.  It would be even more complicated if the X-ray heating transition also occurs during reionization (which it could if we calibrate to the local X-ray luminosity-star formation rate relationship).  On the other hand, they could easily have much larger X-ray luminosities, perhaps eliminating the absorption epoch altogether.

A substantial quasar population (seeded by the remnants of massive Pop III stars, for example) could also eliminate the absorption epoch, because such objects have large X-ray luminosities.  We have conservatively ignored this possibility because the known population of bright quasars contributes little to reionization \citep{fan04} and does not affect the thermal history significantly \citep{venkatesan01}, but this is not to say that ``miniquasars" will be unimportant \citep{oh01, glover03, madau04, kuhlen05-sim} -- although they are unlikely to dominate the ionizing photon budget \citep{dijkstra04, salvaterra05-xray}.  These sources would only affect our predictions before X-ray heating from stars becomes significant -- thus observing the absorption epoch (or lack thereof) could help to constrain their significance.  At lower redshifts, and especially during reionization, miniquasars are unimportant for the 21 cm background because stars easily suffice to heat the IGM above $T_\gamma$.

Another uncertainty is our simple model for shock heating \citep{furl04-sh}.  While it appears to provide a reasonable description of shocks produced as part of the cosmic web \citep{kuhlen06-21cm}, some simulations may imply that the low-density IGM could also host weak shocks \citep{gnedin04}.  These are not included in the model and deserve further study.  As with quasars, these would not affect our predictions during reionization, but they could shorten or even eliminate the absorption epoch.  

Clearly, measuring the mean $\dtb(z)$ would offer insight into the earliest generation of luminous sources \citep{madau97,shaver99,tozzi00,gnedin04}.  How useful might these measurements be for quantitative constraints?  \citet{sethi05} computed the sky-averaged 21 cm signal, including \lya coupling, X-ray heating, and ionization in a restricted range of models (including non-uniform X-ray heating but ignoring \lyn photons, Pop III stars, and feedback, for example).  The qualitative features are similar to our Pop II models, although we predict a much wider range of possibilities given the uncertainties in the models.  

\begin{figure}
\begin{center}
\resizebox{8cm}{!}{\includegraphics{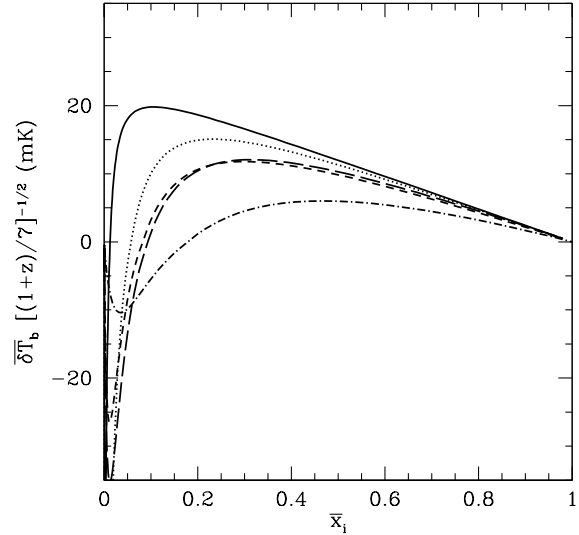}}\\%
\end{center}
%\plotone{dt-xion.eps}
\caption{Brightness temperature as a function of $\bxion$ in several different scenarios.  We have removed the redshift dependence of the underlying signal for clarity.  The solid and short-dashed curves show our fiducial Pop II and Pop III cases, respectively.  The dotted curve uses sets $f_X=0.2$ for Pop II stars, and the dot-dashed curve uses $\fesc=1$ and $f_X=0.5$ for Pop III stars.  Finally, the long-dashed curve corresponds to the solid curve in Fig.~\ref{fig:poptrans-glob}.}
\label{fig:dt-xion}
\end{figure}

Treating the efficiency of each process as a free parameter, \citet{sethi05} considered how well these parameters could be measured from $\bdtb(z)$.  Because (with Pop II stars) heating generally occurs well before reionization, he found that $\bxion(z)$ could be well-constrained.  We find considerably larger degeneracies.  Figure~\ref{fig:dt-xion} shows the brightness temperature as a function of $\bxion$, with the expected redshift dependence of $\dtb$ in a fully neutral universe removed.  We show several different scenarios with a wide range of ionizing and X-ray heating efficiencies. To determine $\bxion$ precisely, we would hope that all these curves would overlap.  Indeed they do approach each other for $\bxion \ga 0.5$, but earlier in reionization the histories differ substantially.  In most cases, the $(1-T_S/T_\gamma)$ factor has not yet saturated in this regime, either because of delayed \lya coupling (usually the case for Pop III stars) or relatively late X-ray heating (particularly in the dotted curve).  Thus we predict that, in the absence of other constraints, $\bxion$ cannot be determined with certainty from the 21 cm background alone unless it is large.

%%%%%%%%%%%%% FIGURE 4: dt v. xion

\begin{figure}
\begin{center}
\resizebox{8cm}{!}{\includegraphics{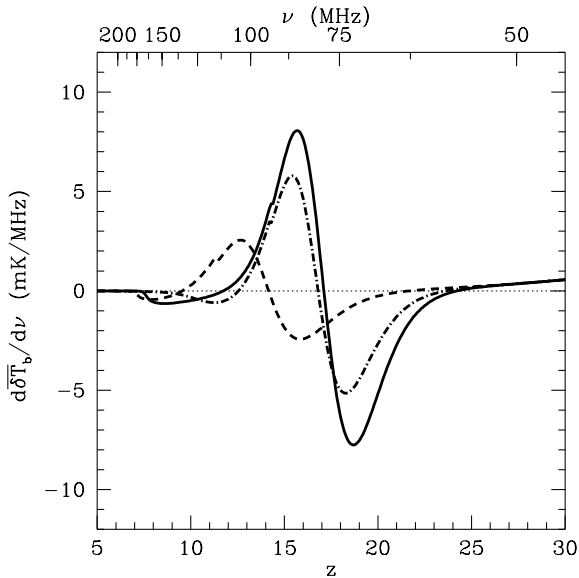}}\\%
\end{center}
%\plotone{gradplot.eps}
\caption{Gradient in the cosmological 21 cm signal for several models:  our fiducial Pop II history (solid curve), our fiducial Pop III history (dashed curve), and a history with strong feedback (dot-dashed curve, corresponding to the solid curve in Fig.~\ref{fig:poptrans-glob}).  The small hitches in each curve occur at $T_S \approx T_\gamma$.}
\label{fig:gradplot}
\end{figure}

The most straightforward observation would be to measure $\dtb(z)$ directly.  Because this is an all-sky signal, single-dish measurements (even with a modest-sized telescope) can easily reach the required mK sensitivity.  However, the much stronger synchrotron foregrounds (primarily from the Galaxy; see \citealt{furl06-review} for a review) nevertheless make such observations extremely difficult.  As a rule of thumb, a ``quiet" portion of the sky has a brightness temperature
\begin{equation}
T_{\rm sky} \sim 180 \ \left( \frac{\nu}{180 \MHz} \right)^{-2.6} \kel.
\label{eq:tsky}
\end{equation}
Clearly, the amplitude of the cosmological signal will be impossible to extract from the foregrounds.  But the spectral structure of the 21 cm background imprints features on the (otherwise smooth) spectrum that may be separable.   Figure~\ref{fig:gradplot} shows the expected gradients in a few of our models.  During reionization, we typically have $\deriv \bdtb/\deriv z \la 1 \mkel \MHz^{-1}$, over three orders of magnitude smaller than the foregrounds.  Without feedback, the gradients tend to increase throughout reionization; with feedback, they peak earlier on.  The gradients can be much larger during the absorption era, partly because the signal can vary over a much wider range and partly because $\deriv \nu/\deriv z$ also decreases.  Of course, the foreground gradients also increase significantly at these low frequencies (by about an order of magnitude at $\nu \approx 90 \MHz$ according to eq. \ref{eq:tsky}), so it is not clear which regime will actually be easier to constrain.  Separating the cosmological signal promises to be a challenge \citep{shaver99, gnedin04} and will be easier in histories with strong features (in which feedback does not slow reionization, for example).

%%%%%%%%%%%%% FIGURE 5: Gradients

The limiting factors in direct measurements are likely to be systematics \citep{shaver99}.  Foregrounds vary across the sky, which couples to the frequency dependent beam and sidelobes (though the variation should be smooth and hence removable; \citealt{oh03}).  One strategy to avoid confusion with features in the foreground is to use multiple (wide) fields, in which foregrounds are presumably uncorrelated but across which the cosmological signal would be constant.  Hydrogen recombination lines leave spectral features throughout the relevant frequency range, but they can be removed through follow-up observations with high spectral resolution.  Finally, the biggest challenge will be calibrating the telescope response to better than one part in $10^5$ over a wide frequency range.  Internal loads of sufficient quality may be difficult to find.  

Fortunately, even if systematics do prove to be insurmountable, there are other ways to extract the mean signal.  \citet{cooray05-glob} pointed out that scattering by a massive, low-redshift galaxy cluster can shift the spectrum in frequency space, so that differential measurements between the cluster and background may reveal the global spectrum.  This avoids many systematics, but given the relatively small angular size of clusters may not measure the truly ``global" signal.  A second possibility is to use the 21 cm fluctuation amplitude to measure the mean background, provided three-dimensional anisotropies can be measured \citep{barkana05-vel}.  For example, \citet{mcquinn05-param} showed that, when density (and velocity) fluctuations dominate the signal, the overall signal amplitude can be extracted by combining 21 cm data with CMB data.  (They phrased their constraint as $\bxion$, but in more general models it corresponds to the mean brightness temperature.)

Despite these difficulties, the global 21 cm background would be an extremely useful tool in understanding the overall evolution of the first sources of light.  Together with the fluctuating signal (see \citealt{furl06-review} and references therein), the global background could reveal many of the important properties of the ``twilight zone" of structure formation.  

I thank S. P. Oh for insightful comments on the manuscript, A. Loeb and J. Pritchard for helpful discussions, and R. Barkana for sharing the stellar spectra used in \S \ref{tshist}.  I also thank the Tapir group at Caltech for hospitality while some of this work was completed.

%\bibliographystyle{mn2e}
%\bibliography{Ref_21cm}

\end{document}